# On reconstruction of chaotic attractor from time series represented as "clusters"

A.A.KIPCHATOV and L.V.KRASICHKOV

College of Applied Science, Saratov State University,

83 Astrakhanskaya, Saratov 410071, Russia

**Abstract** - The method of reconstruction of an attractor from a set of short time series (*clusters*) is proposed and discussed. This method is most useful for correlation dimension estimation of experimental data.

## INTRODUCTION

Time delay method proposed by Packard et al. [1] and Takens [2] and the correlation dimension algorithm of Grassberger and Procaccia [3] are main tools for characterizing of dynamical systems from scalar time series [4]. However, to reconstruct the underlying attractor of oscillation and especially to estimate its dimension, the long enough time series are required [5–7]. At the same time, there are many experiments where short time series can be considered only, and hence the methods of evaluation of the dimension from small data sets are of principal interest [8,9].

In this paper we propose the method of attractor reconstruction of dynamical system from a set of short time series and show the validity of the method from correlation dimension estimations.

## BASIS OF THE METHOD

The attractor of smooth $m$–dimensional dynamical system can be reconstructed in the embedding space from Takens' theorem [2]. Let $u(t_k)$ be a scalar time series measured from continuous signal $u(t)$ with the sampling time $T_s$ ($t_k = kT_s$), where $k = 0, 1, ..., N_d$ and $N_d$ is the number of data points in the time series. Then the state vectors in the reconstructed $d_E$–dimensional embedding space are defined by

$$\mathbf{v}_n \equiv \mathbf{v}(t_n) = \{u(t_n), u(t_n + \tau), ..., u(t_n + \tau(d_E - 1))\}, \qquad (1)$$

where $n = 1, ..., N_d - l(d_E - 1)$ and $\tau$ is the delay time ($\tau = lT_s, l = 0, 1, ...$). If $d_E \geq 2m + 1$, then properties of attractor in original phase space as well as in embedding space are equal.

Note that Takens' theorem requires the infinite number of points in continuous time series and the infinite resolution of data. In this case, the embedding dimension $d_E$ is unique parameter of the method (1). Clearly in practice any data of experimental systems do not satisfy the above requirements, and hence the appropriate choice of the delay time $\tau$ is of importance also. The correct choice of $\tau$ and $d_E$ allows to use the shorter realizations in order to calculate dimension and such choice is not trivial (see, for example, [10] and references therein). However, even in the case of proper reconstruction of an attractor from (1), the calculations of correlation dimension of attractor require the number of points $N_{min}$ (e.g., see [5–7]). It is known that the larger the dimension to be estimated, the larger the needed $N_{min}$. Real experimental data have not the continuous regions of $N_{min}$ points often and can not provide the above requirements. However, it is possible to obtain a set of time series of same process measured at various times and such data can be successfully used also to reconstruct attractor by the time delay method.

The basic idea of our method is in accumulation a large enough number of points in reconstructed attractor using a set of short time series instead of the long one. Thus, the short time series must provide the reconstruction of only one state vector (1), i.e. length of short time series is not less than $N_{short} = l(d_E - 1) + 1$ points. The short time series allowing to reconstruct, at least, the one state vector will be called *clusters*. In this case, in order to reconstruct attractor in embedding space we must have $N_L$ of such time series. The Takens' method suggests that the continuous set of the embedding vectors $\mathbf{v}_n$ is yielded from the time series $u(t_k)$, but the continuous phase trajectory is important, e.g. in estimations of Lyapunov exponents. Whereas the dimension is characterized by location of attractor points in the phase space, and does not require of any sequence of appearance of attractor points. If any single point reconstructed from eq.(1) belongs to the attractor, then all points reconstructed from different clusters belong to this attractor too. The dimension of such attractor must be equal to the dimension of the attractor reconstructed from continuous time series. Note that there is no formal contradiction of the reconstruction method proposed here and the Takens' theorem. It is obvious that the portions of time series must be yielded by the same dynamical process.

To test presented method we calculated the correlation dimension [3] using the correlation integral [11–13] is given by

$$C(\varepsilon) \approx \frac{1}{M} \sum_{i=1}^{M} \frac{1}{N} \sum_{j=1}^{N} \theta(\varepsilon - \| \mathbf{v}_i - \mathbf{v}_j \|), \qquad (2)$$

where $\varepsilon$ is the scaling length, $M$ is the number of reference points, $N$ is the number of data points, $\theta$ is the Hevisaide function and the vertical bars denote the norm of the vector.

The correlation dimension $D_c(\varepsilon)$ is calculated from the following equation

$$D_c(\varepsilon) = \frac{\log(C(\varepsilon)/C(\varepsilon + \delta))}{\log(\varepsilon/(\varepsilon + \delta))}, \qquad (3)$$

where $\delta$ is the localization range of correlation function in which the linear least-square fit is performed. We present the results of estimation of the correlation dimension $D_c$ as the function of the length scale $E$ in dB units (i.e. $E = 20 \cdot \log_{10}(\frac{\varepsilon}{\varepsilon_0})$, where $\varepsilon_0$ is the total extantion of attractor) and fix the localization range $\Delta = 20 \cdot \log_{10}(\frac{\delta}{\varepsilon_0}) = -50$dB.

## IMPLEMENTATION OF PROPOSED METHOD

In our tests the suggested above method we used the time series of $x(t)$ variable of the Rössler system [14] given by

$$\dot{x} = -(y+z), \dot{y} = x + 0.2y, \dot{z} = 0.2 - 4.6z + xz. \qquad (4)$$

In order to obtain the data set $x(t)$, the eqs.(4) were integrated using the fourth-order Runge-Kutta algorithm with the time step $dt = 0.04$ (the sampling time $T_s = 5dt = 0.2$). The first $2 \cdot 10^4$ iterates were ignored as transients for all results presented below.

The correlation dimension $D_c(E)$ of the attractor reconstructed from long continuous time series is shown in fig.1 (dotted curve) for further comparisons with our method. According to Takens' theorem, $d_E = 7$ was chosen. Note that we used same values of $\tau$, $d_E$, $N$ and $M$ (see eq.(2)) for all results throughout this paper, i.e. $N$ points of reconstructed attractor were accumulated from clusters.

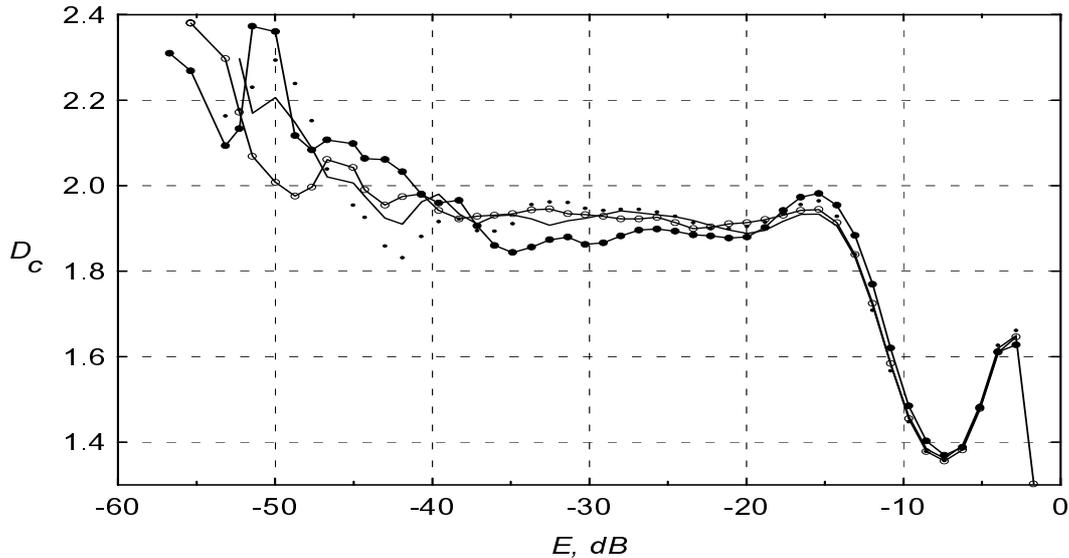

Fig. 1. The correlation dimension $D_c(E)$ are estimated from the long continuous time series (dotted curve) and from sets of clusters. All results were obtained for $x(t)$ of the Rössler system eqs.(4) ($\tau = 5T_s = 1.0$, $d_E = 7$, $N = 2 \cdot 10^4$, $M = 10^3$).

Consider three examples where the attractor can be reconstructed and its correlation dimension can be estimated from short data sets.

In the first example, we used the short time series which allowed to reconstruct one embedding vector only (i.e., $N_{short} = 31$) and the time separation between the time series $N_{ts}$ points was chosen from a standard random process with uniform distribution (the $N_{ts}$ was set in interval $[1, 512]$). In the case, the total number of embedding vectors $N = N_L = 20000$ was accumulated. The correlation dimension $D_c(E)$, for this case, is shown in fig.1 (solid curve).

In the second example, the set of clusters with the length of each cluster in the set as well as the time separation of the series were generated on the basis of random process ($N_{short} \in [31, 95]$, i.e. number of the embedding vectors from 1 to 65 and $N_{ts} \in [1, 512]$), as in the above case. Fig.1 (solid curve with circles) shows the result of calculation of correlation dimension.

Finally, in the third experiment, we consider the dynamical process which is generated during the short time interval and this interval does not allow to obtain $N_{min}$ points in time series. If it is possible to repeat the process over and over, then our method is also performed. Such situation was modelled by accumulation of the attractor points from short time series ($N_{short} = 54$, i.e. number of the embedding vectors is equal to 24). The number of used clusters, in this case, was equal to $N_L = 835$. Each time series is generated for new initial conditions (namely, $x, y, z$ in eqs.(4) were selected randomly in interval $[-10, 10]$ with step 0.005), after transient process (first 20000 points). Result is shown in Fig.1 (solid curve with disks).

The estimations of the correlation dimension have confirmed the similarity of structures of the attractors for wide range of the length scales and thus the validity of prorosed method.

## CONCLUSION

The proposed method of attractor reconstruction from nontotal (broken) data sets can be very useful in many experimental investigations. This method can be effectively applied to characteristics estimating of processes presented by short and fragmented data (e.g., from economical, climatical and other systems). Such reconstruction will allow to investigate the complicated oscillations of high-frequency systems and processes (i.e. the cases when we come across difficulties because of the pulse nature of the process and since it is impossible to obtain the long continuous time series on the basis of direct analog-to-digital (AD) conversion). The third example shows the way to work with pulse time series. The high-frequency AD conversion can be performed, for instance, by using the storage oscilloscope and through further accumulation of obtained data.

Another useful effect of proposed reconstruction from a set of short time series (clusters) lies in more uniform filling of attractor by points. This effect leads to minimization of statistical errors of the Grassberger-Procaccia algorithm when finite data sets are used [15].

*Acknowledgments* – This work was supported by The Russian Fund of Fundamental Research under Grant No. 93-02-16171.